\documentclass[preprint,3p,11pt,twocolumn
]{elsarticle}

\usepackage{graphicx}% Include figure files
\usepackage{amsmath} 

\usepackage{amssymb}
\usepackage{dcolumn}% Align table columns on decimal point
\usepackage{bm}% bold math
\usepackage[dvipsnames]{xcolor}
\usepackage{mathtools, cuted}
\DeclareMathOperator\erf{erf}

\begin{document}
\begin{frontmatter}		
\title{Modeling Stratified Segregation in Periodically Driven Granular Heap Flow}
%...unsteady periodic granular heap flow
%...modulated granular heap flow
%...periodic granular flow on a heap
%...in periodically forced flow on a granular heap
%...in periodically driven granular heap flow

\author[1,4]{Hongyi Xiao}

\author[2]{Zhekai Deng}

\author[1,2,3]{Julio M. Ottino}

\author[1]{Paul B. Umbanhowar}

\author[1,2,3]{Richard M. Lueptow}
\ead{r-lueptow@northwestern.edu}

\address[1]{Department of Mechanical Engineering, Northwestern University, Evanston, IL 60208, USA}
\address[2]{Department of Chemical and Biological Engineering, Northwestern University, Evanston, IL 60208, USA}
\address[3]{The Northwestern University Institute on Complex Systems (NICO), Northwestern University, Evanston, IL 60208, USA}
\address[4]{Department of Physics and Astronomy, University of Pennsylvania, Philadelphia, PA 19104, USA}

\date{\today}

\begin{abstract}
We present a continuum approach to model segregation of size-bidisperse granular materials in unsteady bounded heap flow as a prototype for modeling segregation in other time varying flows. In experiments, a periodically modulated feed rate produces stratified segregation like that which occurs due to intermittent avalanching, except with greater layer-uniformity and higher average feed rates.  Using an advection-diffusion-segregation equation and characterizing transient changes in deposition and erosion after a feed rate change, we model stratification for varying feed rates and periods. Feed rate modulation in heap flows can create well-segregated layers, which effectively mix the deposited material normal to the free surface at lengths greater than the combined layer-thickness.  This mitigates the strong streamwise segregation that would otherwise occur at larger particle-size ratios and equivalent steady feed rates and can significantly reduce concentration variation during hopper discharge. Coupling segregation, deposition and erosion is challenging but has many potential applications. 
\end{abstract}

\begin{keyword}
segregation \sep heap flow \sep continuum modeling \sep hopper discharge \sep unsteady flow
\end{keyword}

\end{frontmatter}

\section{Introduction}
\label{intro}
Flowing granular materials spontaneously separate due to particle size or density differences~\cite{Savage1988, Gray2018, umbanhowar2019} to form a variety of segregation patterns~\cite{Ristow2000, Aranson2006, Meier2007}. The degree of segregation and the nature of the segregation pattern depend on the complex interactions between the flow dynamics and the segregation dynamics~\cite{Meier2007, Makse1991, Gray2009, Gray2005c}. A successful approach to predicting species segregation in flowing granular materials is based on a modification of the continuum advection-diffusion equation that includes a term accounting for segregation of each species~\cite{umbanhowar2019}. This approach works well for steady segregating flows including heaps~\cite{Fan2014a, Schlick2015c, deng2019modeling, HongyiDensity, drahunBridgwater1983, Isner2020a}, inclined planes~\cite{Deng2018,Dolgunin1998, Wiederseiner2011a, Marks2012, Gray2006b, Dolgunin1995, Dolgunin1998, Gray2005c}, and wall-driven flows~\cite{fry2019}, as well as for transient segregation in rotating tumblers~\cite{Schlick2015b, deng2019modeling, WassgrenLiu2019} and unsteady flows in complicated geometries like hoppers~\cite{xiao2019continuum, WassgrenLiu2019, Deng2020}. Here we use this continuum advection-diffusion-segregation model to better understand the even more complex interaction between flow dynamics and segregation dynamics that results in stratified layers of segregated particles for time-periodic heap flow~\cite{Xiao2017}.

We consider single-sided quasi-two-dimensional (quasi-2D) bounded heap flows in which particles fall onto the highest point of the heap and flow down the sloped surface until they reach a bounding downstream vertical wall~\cite{fan2017segregation}. For steady feed rates of size-bidisperse particle mixtures, three segregation conditions can result~\cite{Fan2012}.  At low feed rates, intermittent avalanches produce an irregular stratified segregation pattern composed of large and small particle layers that vary in thickness and streamwise extent~\cite{Makse1997, Williams1976, Gray1997, Baxter1998, Fan2012}. At moderate feed rates the downslope flow is steady, and smaller particles fall between larger particles to deposit mainly on the upstream portion of the heap, while larger particles rise to the top of the flowing layer and, consequently, deposit mostly on the downstream portion of the heap~\cite{Fan2012}.  At sufficiently high feed rates, particles have too little time to segregate and deposit on the heap in a nearly fully mixed condition~\cite{Fan2012,Fan2014a}.  

In many practical situations in industry, a fully mixed condition is preferred but difficult to achieve because of the high feed rates that are required. Alternative methods to prevent segregation include altering particle characteristics such as density~\cite{Duan2021, Tunuguntla2014, gray_ancey_2015, Jain2005, Jain2005b, Felix2004, Larcher2015} or elasticity~\cite{Brito2008}, or adding small amounts of liquid to dry particle mixtures to make them cohesive~\cite{Li2003, Samadani2000, Samadani2001, Liao2010, Liu2013}. However, these approaches are not always feasible or appropriate in application. Here we focus on an alternative approach, that of unsteady flow of a size-bidisperse mixture of particles on a heap created by periodic feed-rate modulation, which results in a regular segregation pattern of stratified layers of small and large particles that is ``mixed" at length scales larger than the stratification layer thickness~\cite{Xiao2017}. 

Unlike a low constant feed rate, which typically results in quasi-periodic avalanches that lead to irregularly stratified layers~\cite{Makse1997, Williams1976, Gray1997, Baxter1998, Fan2012}, the modulated feed rate approach considered here allows moderate cycle-averaged feed rates without the usual segregation of small particles in the upstream region and large particles in the downstream region that occurs for equivalent steady feed rates or the uneven stratification that occurs at low feed rates~\cite{Xiao2017}. Consider, for example, filling a hopper with a strongly segregating mixture of small and large particles in an industrial situation where the preference is for the particles to remain mixed to assure product uniformity. Although this can be achieved using high feed rates~\cite{Fan2012}, such high feed rates can be difficult and costly to achieve in practice. Instead, modulating the feed rate to force a regular stratified pattern of small and large particle layers results in greater overall uniformity and significantly reduced segregation upon discharge~\cite{Xiao2017} at low to moderate average feed rates. In addition to its practical application, the stratification pattern formation is driven by a strong coupling between segregation and erosion and deposition~\cite{Xiao2017,Xiao2017_2,Lueptow2017}, which is a combination of processes that is not well understood or modelled.

Our approach to investigating the effects of a modulated feed rate on the segregation of size-bidisperse mixtures in bounded heap flows is to implement the unsteady form of a continuum advection-diffusion-segregation model~\cite{umbanhowar2019} combined with a model of the transient heap flow kinematics when the feed rate is changed~\cite{Xiao2017_2}. This permits analysis of the interaction of the transient flow with the segregation to better understand the origin of the stratified structures of small and large particles observed in experiments~\cite{Xiao2017} and to predict the impact of feed rate modulation parameters on the resulting stratification patterns.

\begin{figure}[t]
	\centering
	\includegraphics[width=0.9\linewidth]{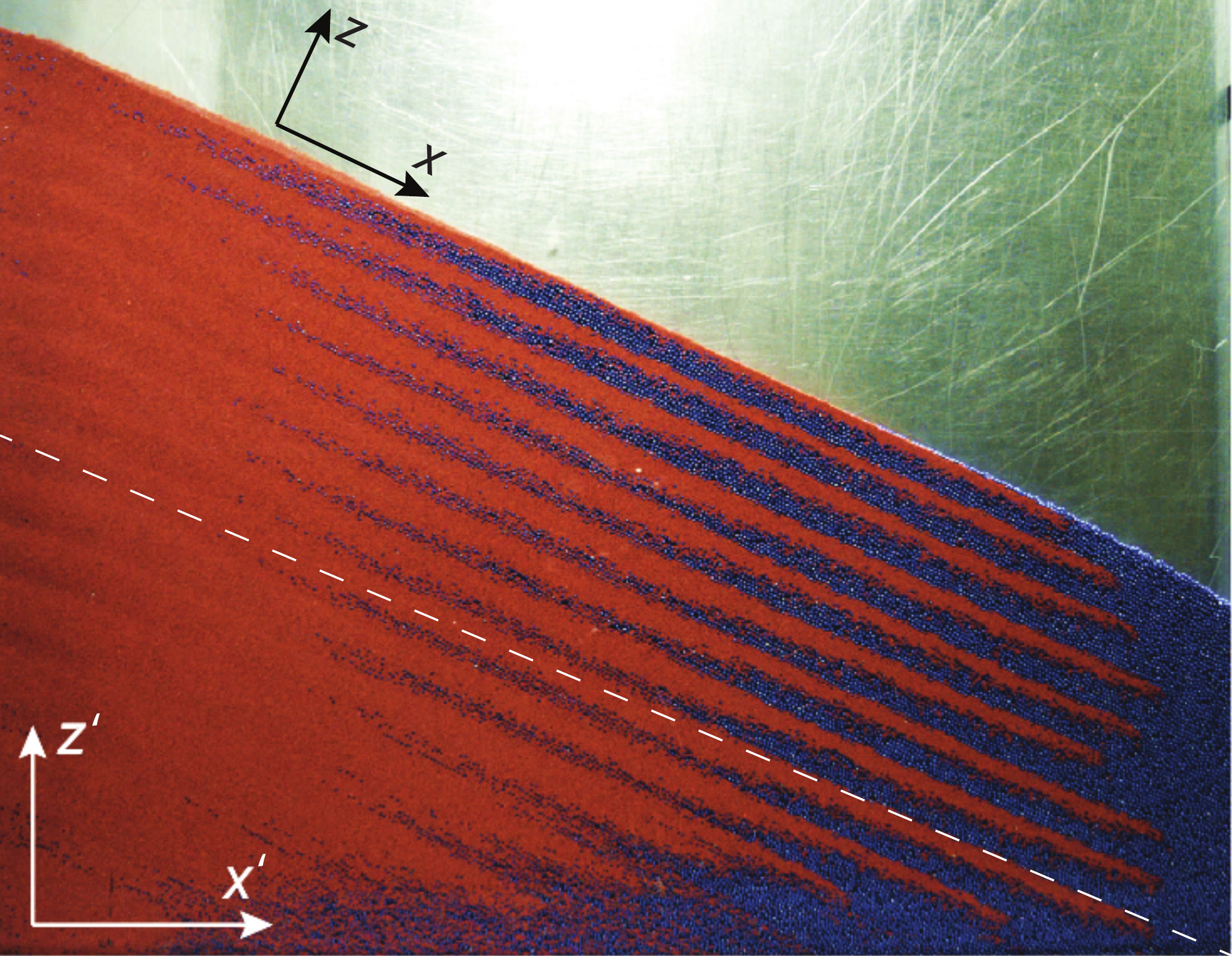}
	\caption[Stratified segregation pattern generated by a modulated feed rate of a size-bidisperse mixture]{Stratified segregation pattern (digitally enhanced contrast) of a size bidisperse mixture ($0.5$\,mm diameter red particles and $2$\,mm diameter blue particles) fed onto a quasi-2D bounded heap with width $W = 69$\,cm and thickness $T = 1.27$\,cm by alternating between a fast flow rate of $q_\mathrm{f} =37.0 $\,cm$^2$/s for $t_\mathrm{f} =3$\,s and a slow flow rate of $q_\mathrm{s}=0.4$\,cm$^2$/s for $t_\mathrm{s}= 62$\,s~\cite{Xiao2017}. Dashed line indicates boundary between material deposited during bounded (above) and unbounded (below) heap flow; here we study the former.}
	\label{Stratified_pattern_exp}
\end{figure}

\section{Background and the Continuum Segregation Model}\label{sec:segmodel}
The specific problem that we consider here is the formation of stratified segregation patterns in a quasi-2D single-sided bounded heap driven with a modulated feed rate, an example of which is shown in Fig.~\ref{Stratified_pattern_exp}~\cite{Xiao2017}. In experiments, a well-mixed size-bidisperse equal-volume mixture of small (red, $d_s=0.5$\,mm) and large (blue, $d_l=2$\,mm) spherical glass particles is fed from a screw feeder at a volumetric feed rate, $Q,$ into the $T= 12.7$\,mm gap between parallel vertical plates having a horizontal width of $W=69$\,cm~\cite{Xiao2017}. The particles segregate as they flow in a thin layer down the $L$ long slope from the feed location to the downstream bounding wall. To achieve stratified patterns of large and small particle layers, $Q$ is modulated by alternating between a faster and a slower feed rate. The resulting 2D feed rates at the upstream end of the flowing layer, $q_j = Q_j/T$, are applied for times $t_j$, and produce flowing layer thicknesses $\delta_j$, where $j\in\{\mathrm{f,s}\}$ for fast and slow, respectively. Note that the local flow rate on the heap, $q(x,t)$, is maximum at the upstream end of the flowing layer, where it equals the feed rate, i.e.\ $q(x=0,t) = q_j$, and decreases linearly to zero at the downstream endwall during steady flow as particles are deposited uniformly onto the heap.

\citet{Xiao2017} studied segregation patterns in one-sided quasi-2D heap flow experiments like the example shown in Fig.~\ref{Stratified_pattern_exp} by varying the flow rate modulation parameters $q_j$ and $t_j.$ In that work, the stratification patterns were postulated to be primarily driven by the time-varying flow kinematics, specifically the propagation and relaxation of the local flow rate on the heap from the feed zone to the downstream end wall, and the associated particle segregation. 

What needs to be done to better understand the interaction of time-varying flow and segregation is to extend the application of a continuum segregation model to accommodate the modulated feed rate and subsequent variation in flow down the heap. This model is based on the advection-diffusion equation with an additional term for segregation~\cite{drahunBridgwater1983, Bridgwater1985}. It has been successfully applied for modeling steady, developing, and transient granular flows of size-segregating or density-segregating materials~\cite{Gray2006b, Thornton2006, Fan2014a, Tunuguntla2014, HongyiDensity, deng2019modeling, Gray2018, umbanhowar2019, Duan2021}. We use a form of the transport equation incorporating a relatively simple expression for the segregation term~\cite{Fan2014a, umbanhowar2019}, specified as:
\begin{strip}
\begin{equation}
\frac{\partial c_i}{\partial t} + \underbrace{ \frac{\partial (u c_i)}{\partial x} + \frac{\partial (w c_i)}{\partial z}}_\text{advection} +  \underbrace{\frac{\partial (w_{p,i}c_i)}{\partial z}}_\text{segregation} - \underbrace{ \frac{\partial }{\partial x}(D \frac{\partial c_i}{\partial x}) + \frac{\partial}{\partial z}(D \frac{\partial c_i}{\partial z})}_\text{diffusion} = 0,
\label{bidisperse_eqn}
\end{equation}
\end{strip}

\noindent where $c_i$ is the local concentration of species $i$, $u$ and $w$ are the mean granular velocity in the streamwise ($x$) and surface-normal ($z$) directions, respectively, and $D$ is the collisional diffusion coefficient. The time rate of change of species concentration $c_i$ (1st term) depends on advection due to mean flow (2nd and 3rd terms), segregation via a mixture-specific percolation velocity for each of the species, $w_{p,i} = w_i - w$ (4th term), where $w_i$ is the local velocity component for species $i$ and $w$ is the local velocity component of the mixture normal to the free surface, and collisional diffusion (last two terms). 

A key aspect of this model is the dependence of the percolation velocity of species $i$ on the local shear rate and the local concentration of the other particle species~\cite{Fan2014a,Schlick2015c} such that $w_{p,i} = S \dot{\gamma}(1-c_i)$, where $S=S_R d_s\ln{d_l/d_s}$ is a mixture-specific segregation length scale parameter and $\dot{\gamma} = \frac{\partial u}{\partial z}$ is the local shear rate. The diffusion coefficient, $D,$ is calculated as $D=C_D \dot{\gamma} \bar{d}^2$, where $C_D$ is a material dependent property and $\bar{d}$ is the average particle diameter~\cite{Utter2004, tripathi_khakhar_2013, Fan2015a}.  Here we use $S_R=0.26$ and $C_D=0.1$, consistent with previous studies~\cite{xiao2019continuum, Deng2020}, although results are not strongly dependent on either parameter's exact value~\cite{Fan2014a}. We further assume, as usual, that the streamwise diffusion term in Eq.~\ref{bidisperse_eqn} is negligible as a consequence of $\delta_j/L \ll 1$~\cite{umbanhowar2019}.

Equation~\ref{bidisperse_eqn} can be solved numerically for the local concentration of the two particle species at all points in the flow for the steady case using an operator splitting approach~\cite{Fan2014a,Schlick2013}.
However, implementing a modulated flow rate in the continuum segregation model is more challenging because the local flow rate, flowing layer thickness, and, consequently, velocity field can all vary continuously with respect to time and streamwise location depending on the modulation parameters. In our previous ``instantaneous flow transition" modelling approach for this problem~\cite{Lueptow2017}, we simply ignored the unsteady kinematics by applying the model using fully-developed steady flow conditions~\cite{Fan2014a} and instantaneously changing the velocity field and flowing layer thickness everywhere in the flowing layer from that corresponding to a low feed rate to that for a high feed rate, and vice versa. Thus, Eq.~\ref{bidisperse_eqn} is solved at one flow rate as a steady flow, say at the slow flow rate $q_\mathrm{s}$ with the value for $\delta_\mathrm{s}$ based on a steady-state empirical correlation~\cite{Schlick2015c} for duration $t_\mathrm{s}$. Then the conditions are instantaneously changed to those for the fast flow rate $q_\mathrm{f}$, and Eq.~\ref{bidisperse_eqn} is implemented for $q_\mathrm{f}$ as a steady flow with the associated $\delta_\mathrm{f}$ for duration $t_\mathrm{f}$. Although the downstream propagation of the change in flow rate and flowing layer thickness from the feed zone is completely ignored in this approach, repeating this instantaneous modulation of the flow conditions along the entire flowing layer results in a stratified segregation pattern that is qualitatively similar to an experiment at the same conditions~\cite{Lueptow2017}, especially when $t_j$ is large compared to the transient duration associated with the change in feed rate (see next paragraph).  In this paper, we utilize a more sophisticated approach to modeling the modulated flow, described in the next section, that more accurately reflects the transient physics and consequently provides higher fidelity results for the concentration field.

\section{Modeling segregation during feed rate transitions}
\label{sec:trans}

\begin{figure}
	\centering
	\includegraphics[width=0.65\linewidth]{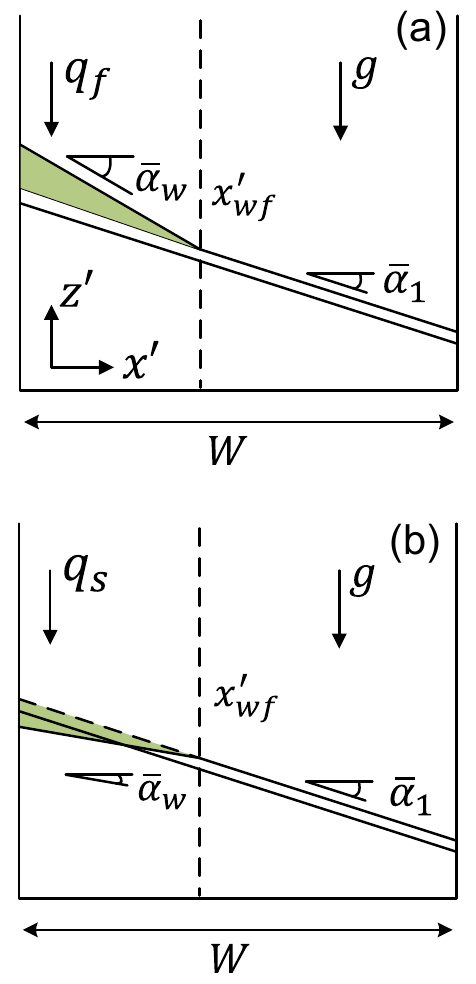}
	\caption[A wedge of material inclined at angle $\bar{\alpha}_\mathrm{w}$ propagates downslope after a step change in the feed rate]{In monodisperse flow, a feed rate change produces a wedge of material inclined at angle $\bar{\alpha}_\mathrm{w}$ that propagates downstream with front location $x^{\prime}_\mathrm{wf}$, while the downstream portion of the heap beyond the wedge remains inclined at angle $\bar{\alpha}_1$ associated with the initial flow rate. (a) A ``positive'' wedge occurs during slow-to-fast transitions (shown shortly after the transition to $q_\mathrm{f}$); (b) a ``negative'' wedge occurs for fast-to-slow transitions (shown shortly after the transition to $q_\mathrm{s}$)~\cite{Xiao2017_2}.}
	\label{Single_Transition}
\end{figure}

When the feed rate of material falling onto the top of a heap is increased, a wedge of material with a steeper angle of repose, $\bar{\alpha}_\mathrm{w}$, forms near the feed zone, as shown in Fig.~\ref{Single_Transition}(a), noting that here we use the horizontal and vertical coordinates, $x^\prime$ and $y^\prime$ with the origin at the lower left corner of the bounding box. Similarly, when the feed rate is decreased, a deficit of material due to bed erosion near the feed zone results in a ``negative" wedge with a smaller surface angle, $\bar{\alpha}_\mathrm{w}$, as shown in Fig.~\ref{Single_Transition}(b). Both wedges propagate downslope as the front of the new flow rate (indicated by $x^{\prime}_\mathrm{wf}$) advances down the heap. Eventually, the wedge propagates to the downstream end of the heap, resulting in a new constant angle of repose along the entire heap surface and steady flow thereafter.  

To accurately model and understand segregation during these transients, consider first the single transitions that occur when the feed rate is changed from slow-to-fast or vice versa. We borrow the kinematics for the transient wedge of bidisperse material associated with a change in feed rate from our previous work on the kinematics for monodisperse feed rate transients~\cite{Xiao2017_2}. Assume that the heap is in steady state at the slow feed rate $q_\mathrm{s}$ when the feed rate is suddenly changed to the fast feed rate $q_\mathrm{f}$, as shown in Fig.~\ref{Single_Transition}(a). After the flow rate is increased to $q_\mathrm{f}$, the surface near the feed zone rises quickly to form a wedge of material with an average surface angle $\bar{\alpha}_\mathrm{w}$ that is steeper than the steady-state average surface angle $\bar{\alpha}_1$ under $q_\mathrm{s}$, while the rest of the heap (downstream of the wedge) continues to rise uniformly with vertical rise velocity $v_\mathrm{r,s} = q_\mathrm{s}/W.$ In time, the wedge front position, $x^{\prime}_\mathrm{wf}$, propagates downstream until it reaches the endwall, after which the entire heap is again in steady state but at a higher repose angle $\bar{\alpha}_2$ associated with $q_\mathrm{f}$. For the opposite case of a fast-to-slow feed-rate transition, Fig.~\ref{Single_Transition}(b), the situation is analogous. After $q_\mathrm{f}$ is reduced to $q_\mathrm{s}$, the rise velocity of the surface near the feed zone decreases quickly and there is a net local outflow of material associated with the reduced repose angle at the slower feed rate. Since the lower portion of the heap continues to rise with velocity $v^{\prime}_\mathrm{r,f}=q_\mathrm{f}/W$, a ``negative" wedge having a lower surface angle $\bar{\alpha}_\mathrm{w}$ forms near the feed zone. The position of the wedge front, $x^{\prime}_\mathrm{wf}$, propagates down the heap as time advances. 

Using mass conservation, we previously proposed a flow model for the front position for monodisperse particles following a sudden change in the feed rate as~\cite{Xiao2017_2}:
\begin{equation}
x^{\prime}_\mathrm{wf} = \sqrt{Ct},
\label{front_position_eqn}
\end{equation}
where $x^{\prime}_\mathrm{wf}$ is the instantaneous wedge front position, $C = \frac{2(q_2 - q_1)}{\tan \bar{\alpha}_\mathrm{w} - \tan \bar{\alpha}_1}$, and $q_1$ and $q_2$ are the feed rates before and after the transition, respectively, which correspond to $q_\mathrm{s}$ and $q_\mathrm{f}$, or vice versa. Using Eq.~\ref{front_position_eqn}, the characteristic time for the wedge front to propagate to the downstream boundary of the flowing layer, i.e.\ $x^{\prime}_\mathrm{wf}=W$, is $\tau=W^2/C$.  Additionally, due to mass balance and the local relation between the flux and surface slope, the transient local flow rate $q(x^{\prime},t)$ down the flowing layer follows a diffusion-like equation~\cite{Xiao2017_2}:
\begin{equation}
\frac{\partial \tilde{q}}{\partial t} = A \frac{\partial^2 \tilde{q}}{\partial x^{\prime 2}},
\label{heap_transition_eqn}
\end{equation}
where $\tilde{q} = q(x^{\prime},t) - q_1 (1-x^{\prime}/W)$, and $A=0.028$\,m$^2$/s is a fitting parameter from the relation, $q=A\tan{\alpha}+B$, between the local flow rate $q$ and the local surface angle $\alpha$, with an additional constant $B=-0.015$\,m$^2$/s~\cite{Xiao2017_2}. (Both $A$ and $B$ are expected to depend on the particle-wall friction coefficient and the ratio of the gap width to the particle diameter.) Solving Eq. \ref{heap_transition_eqn} with appropriate boundary conditions gives~\cite{Xiao2017_2}
\begin{equation}
\tilde{q}(x^{\prime},t) = (q_2 - q_1) \left[1 - \frac{\erf(\frac{x^{\prime}}{\sqrt{4tA}})}{\erf(\frac{x^{\prime}_\mathrm{wf}}{\sqrt{4tA}})}\right],
\label{flow_rate_eqn}
\end{equation}
upstream of the wedge front (i.e.\ in the wedge for $x^\prime < x^{\prime}_\mathrm{wf}$). For regions downstream of the wedge front, $x^\prime > x^{\prime}_\mathrm{wf}$, $q(x^{\prime},t) = q_1\left(1-x^\prime/W\right)$. The corresponding free surface height profile (i.e.\ the ``top" of the flowing layer), $z_\mathrm{t}^{\prime}(x^{\prime},t)$, can be determined via mass conservation $\partial z_\mathrm{t}^{\prime}/\partial t+\partial q/\partial x^{\prime} = 0$. Ideally, the propagation constant $C$ in Eq.~\ref{front_position_eqn} should be specified for each combination of feed rates, which requires extensive calibration. However, since a linear relation exists between $q$ and $\tan{\alpha}$, the value of $C$ should be nearly constant. Here we use $C=3A$ for all simulations for simplicity, which corresponds to $\tan\bar{\alpha}_\mathrm{w}-\tan\bar{\alpha}_\mathrm{1}=\frac{2}{3}(\tan\bar{\alpha}_\mathrm{2}-\tan\bar{\alpha}_\mathrm{1})$, which is an estimate from experimental results~\cite{Xiao2017_2}.

\begin{figure}[t]
	\centering
	\includegraphics[width=\linewidth]{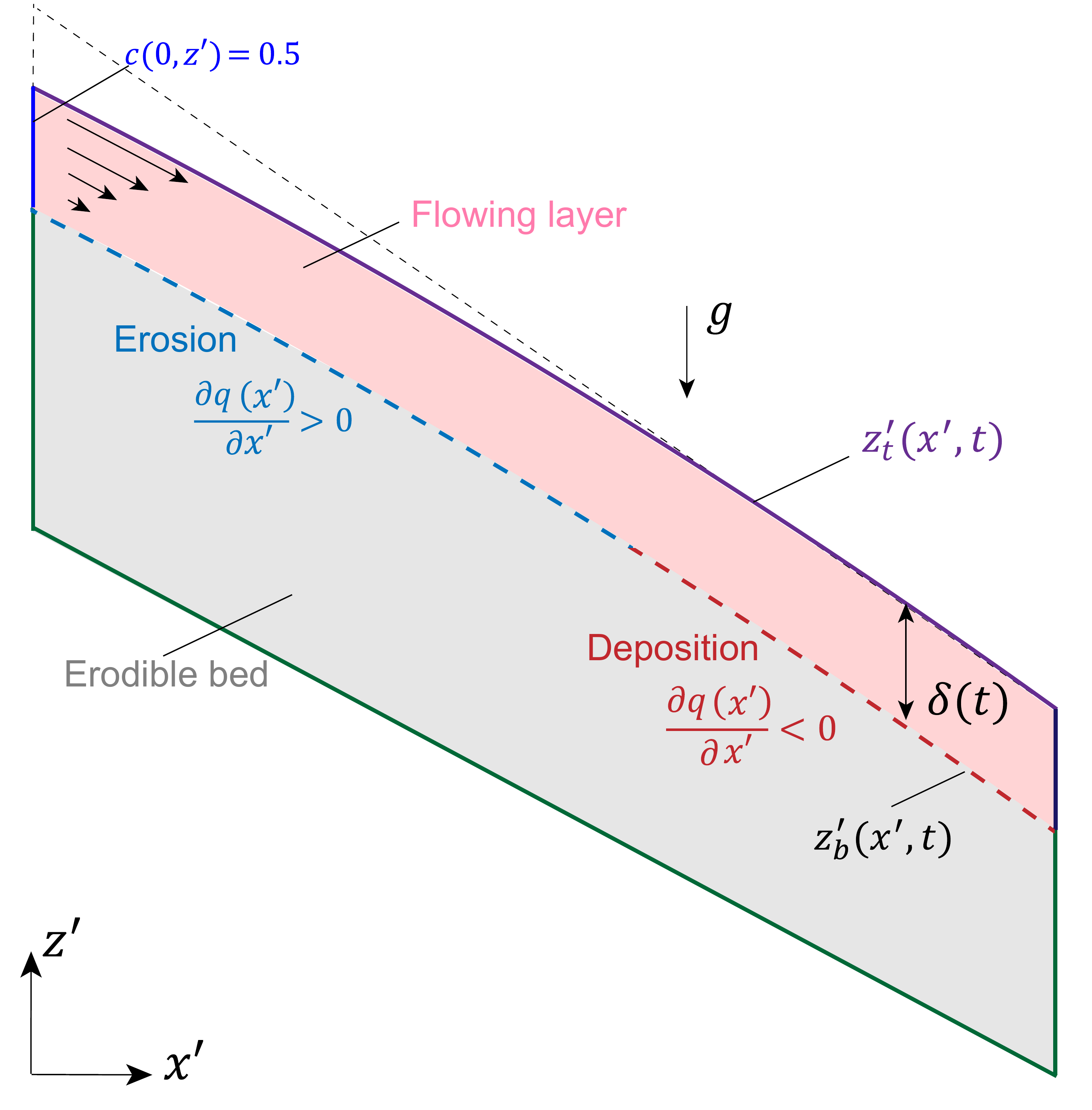}
	\caption{Schematic of the computational domain for the transition from $q_\mathrm{f}$ to $q_\mathrm{s}$, resulting in a negative wedge in the upstream portion of the flowing layer (white region between short-dashed line and flowing layer), where red- and blue-shading indicate the flowing layer and static bed, respectively, and the short-dashed line marks the free-surface location had the feed rate stayed at $q_\mathrm{f}$.}
	\label{setup}
\end{figure}

Assuming that wedge formation kinematics are unaffected by segregation, we apply the monodisperse wedge flow model, Eq.~\ref{flow_rate_eqn}, to segregation in bidisperse flow in conjunction with the continuum segregation model (i.e.,\ Eq.~\ref{bidisperse_eqn}). Because erosion and deposition are important features of the transient segregation process, we model the heap as two coupled regions consisting of a flowing layer and an erodible bed separated by an interface at the bottom of the flowing layer, $z_\mathrm{b}^{\prime}(x^{\prime},t)=z_\mathrm{t}^{\prime}(x^{\prime},t)-\delta(x^{\prime},t)$, see Fig.~\ref{setup}. Deposition on the bed occurs when $z_\mathrm{b}^{\prime}$ increases. The bed concentration, $c_\mathrm{bed}(x^{\prime},z^{\prime},t)$, in the newly deposited region is just the concentration at the bottom of the flowing layer, i.e.\ $c_\mathrm{bed}(x^{\prime},z_\mathrm{b}^{\prime},t)=c(x^{\prime},z_\mathrm{b}^{\prime},t)$. Erosion of the bed occurs when $z_\mathrm{b}^{\prime}$ decreases (the situation shown in Fig.~\ref{setup}), in which case the concentration of material eroded into the base of the flowing layer is  $c(x^{\prime},z_\mathrm{b}^{\prime},t)=c_\mathrm{bed}(x^{\prime},z_\mathrm{b}^{\prime},t)$. In other words, erosion causes particles previously deposited on the bed to be re-incorporated into the flowing layer. This coupling mechanism ensures that the concentration is continuous at the interface between the bed and the flowing layer. The upper and lower boundaries of the flowing layer in Fig.~\ref{setup} are curved due to the coupling of $z'_\mathrm{b}$ and $z'_\mathrm{s}$ to $q$ via Eq.~\ref{heap_transition_eqn}. Further note that from mass conservation and incompressibility, $\partial z'_\mathrm{s}/\partial t + \partial q /\partial x' =0$, erosion occurs when $dq/dx^{\prime}>0$ (corresponding to the situation in the negative wedge for a fast-to-slow transition in Fig.~\ref{setup}) and deposition occurs when $dq/dx^{\prime}<0$ (corresponding to the situation downstream of the wedge front in Fig.~\ref{setup}). 

To simplify the flow kinematics calculation, we assume that the flowing layer thickness is uniform along the entire length of the flowing layer [reducing $\delta(x^{\prime},t)$ to $\delta(t)$], even though it varies slightly with streamwise location and flow rate in experiments~\cite{Fan2014a,Schlick2015c}. The initial flowing layer thickness is set using a previously determined scaling law~\cite{Schlick2015c}: $\delta_j=C_\delta\bar{d}\left[ q_j/\sqrt{g\bar{d}^3}  \right]^{\beta_\delta}$, with $C_\delta=5$, $\bar{d} = (d_\mathrm{l} +d_\mathrm{s})/2$, $\beta_\delta = 0.35$, and $j\in\{\mathrm{f,s}\}$. During a slow-to-fast transition, the flowing layer thickness initially equals $\delta=\delta_\mathrm{s}$ and then increases toward $\delta_\mathrm{f}$ as the front propagates, following $\delta(t) = \frac{x^\prime_\mathrm{wf}}{W} \delta_\mathrm{f} + (1-x^\prime_\mathrm{wf}/W) \delta_\mathrm{s}$, noting that $x^\prime_\mathrm{wf}$ is a function of time (Eq.~\ref{front_position_eqn}). For a fast-to-slow transition, $\delta(t) = \frac{x^\prime_\mathrm{wf}}{W} \delta_\mathrm{s} + (1-x^\prime_\mathrm{wf}/W) \delta_\mathrm{f}$.

The velocity field $(u^{\prime},w^{\prime})$ is required to solve Eq.~\ref{bidisperse_eqn}, but only in the flowing layer since the bed region is static and simply acts as a ``store" of material that is either depleted or supplied by the flowing layer. The velocity field in the flowing layer with $q(x^{\prime},t)$, $z_\mathrm{t}^{\prime}(x^{\prime},t)$, and $z_\mathrm{b}^{\prime}(x^{\prime},t)$ specified must satisfy four requirements. First, the horizontal velocity decreases exponentially with depth~\cite{Fan2014a,xiao2019continuum} while matching the streamwise flux requirement that $q(x^{\prime},t)=\int_{z_\mathrm{b}^{\prime}}^{z_\mathrm{t}^{\prime}} u^{\prime}(x^{\prime},z^{\prime},t)dz$. Second, at the bottom of the flowing layer, $u^{\prime}(x^{\prime},z_\mathrm{b}^{\prime})=0$ and $w^{\prime}(x^{\prime},z_\mathrm{b}^{\prime})=0$, so that the velocity field goes to zero at the fixed bed and is continuous. Third, at the free surface, $w^{\prime}/u^{\prime}=\frac{\partial z_\mathrm{t}^{\prime}}{\partial x^{\prime}}$, ensuring that the flow is always parallel to the free surface. Fourth, the continuity equation, $\partial u^{\prime}/\partial x^{\prime} + \partial w^{\prime}/\partial z^{\prime} = 0$, is satisfied everywhere. The local shear rate, $\dot{\gamma}=\partial u/\partial z$, can be calculated in the coordinates consistent with Eq.~\ref{bidisperse_eqn} by taking the derivative of the streamwise velocity, $u=u^{\prime}\cos{\alpha}-w^{\prime}\sin{\alpha}$, in the perpendicular direction $z$, with $\alpha=-\arctan{(\partial z_\mathrm{t}^{\prime}/\partial x^{\prime})}$. With these constraints realized, the velocity field is~\cite{xiao2019continuum}:
\begin{align}
\begin{split}\label{velocity_labx}
u^{\prime}(x^{\prime},z^{\prime}) = &Mq\left( e^{k(z^\prime-z_\mathrm{t}^\prime)/\delta} - e^{-k} \right),
\end{split}\\
\begin{split}\label{velocity_labz}
w^{\prime}(x^{\prime},z^{\prime}) = &M \frac{\partial q}{\partial x^{\prime}}\left[ (z^{\prime}-z_\mathrm{b}^{\prime})e^{-k}-\frac{\delta}{k}\left( e^{k(z^\prime-z_\mathrm{t}^\prime)/\delta} - e^{-k} \right)\right] \\ &+Mq\frac{\partial z_\mathrm{t}^{\prime}}{\partial x^{\prime}}\left( e^{k(z^\prime-z_\mathrm{t}^\prime)/\delta} - e^{-k} \right),
\end{split}
\end{align}
\noindent where $M=k/\{\delta[1-(1+k)e^{-k}]\}$ and $k=2.3$~\cite{Fan2014a,xiao2019continuum}. In steady state, the velocity field reduces to a simpler form with an exponential profile as described previously~\cite{Fan2014a}.

Having specified the advection field, the segregation term, the diffusion term, and the position of the interface between the bed and the flowing layer, the advection-diffusion-segregation equation (Eq.~\ref{bidisperse_eqn}) is solved numerically using the finite element method~\cite{xiao2019continuum}. The Arbitrary Eulerian Lagrangian (ALE) method handles the moving boundary of the flowing layer and the bed region, and the Streamline Upwind Petrov Galerkin (SUPG) method is used to stably simulate advection dominated regions~\cite{xiao2019continuum} in the flowing layer. In the approximately trapezoidal computational domain shown in Fig. ~\ref{setup}, quadrilateral elements are used with a resolution of 250$\times$80 nodes for the flowing layer and $250\times400$ nodes for the bed, and the integration time step is $0.01$\,s. Results are insensitive to reasonable variations of the grid size and the time-step. In each simulation, the initial condition in the flowing layer is set as the steady state solution to Eq.~\ref{bidisperse_eqn} under the initial feed rate $q_1$, and the bed concentration is initialized using the initial concentration profile at the bottom of the flowing layer, $c(x^{\prime},z_\mathrm{b}^{\prime})$. The simulation parameters match the experiments with $W=69$\,cm, $d_l=2$\,mm, and $d_s=0.5$\,mm. The initial surface profile $z_\mathrm{t}^{\prime}$ is set to linear with a slope of $28^\circ$, which is a simplification based on experimental observation~\cite{Xiao2017,Xiao2017_2}. The slope for the bottom of the computational domain is also set to $28^\circ$ to avoid unnecessary computations for the static heap, and the initial bed depth is set to $0.5q_1/W+0.05$\,m to ensure adequate depth for erosion. As the simulation proceeds, the transient advection field is calculated using Eqs.~\ref{flow_rate_eqn}-\ref{velocity_labz} under the modulation parameters, $q_\mathrm{f}$, $q_\mathrm{s}$, $t_\mathrm{f}$, and $t_\mathrm{s}$.  This field is in turn used to solve Eq.~\ref{bidisperse_eqn} with the boundary movement specified by the calculated positions of the top, $z_\mathrm{t}^{\prime},$ and bottom, $z_\mathrm{b}^{\prime},$ of the flowing layer. This yields the instantaneous flowing layer concentration profile $c(x^\prime, z_\mathrm{b}^\prime \le z^\prime \le z_\mathrm{t}^\prime,t)$ and bed concentration profile $c_\mathrm{bed}(x^\prime,z^\prime \le z_\mathrm{b}^\prime,t)$ through the interface coupling scheme described above.

\section{Model validation}
\label{sec_validation}
\begin{figure*}[t]
	\centering
	\includegraphics[width=0.8\textwidth]{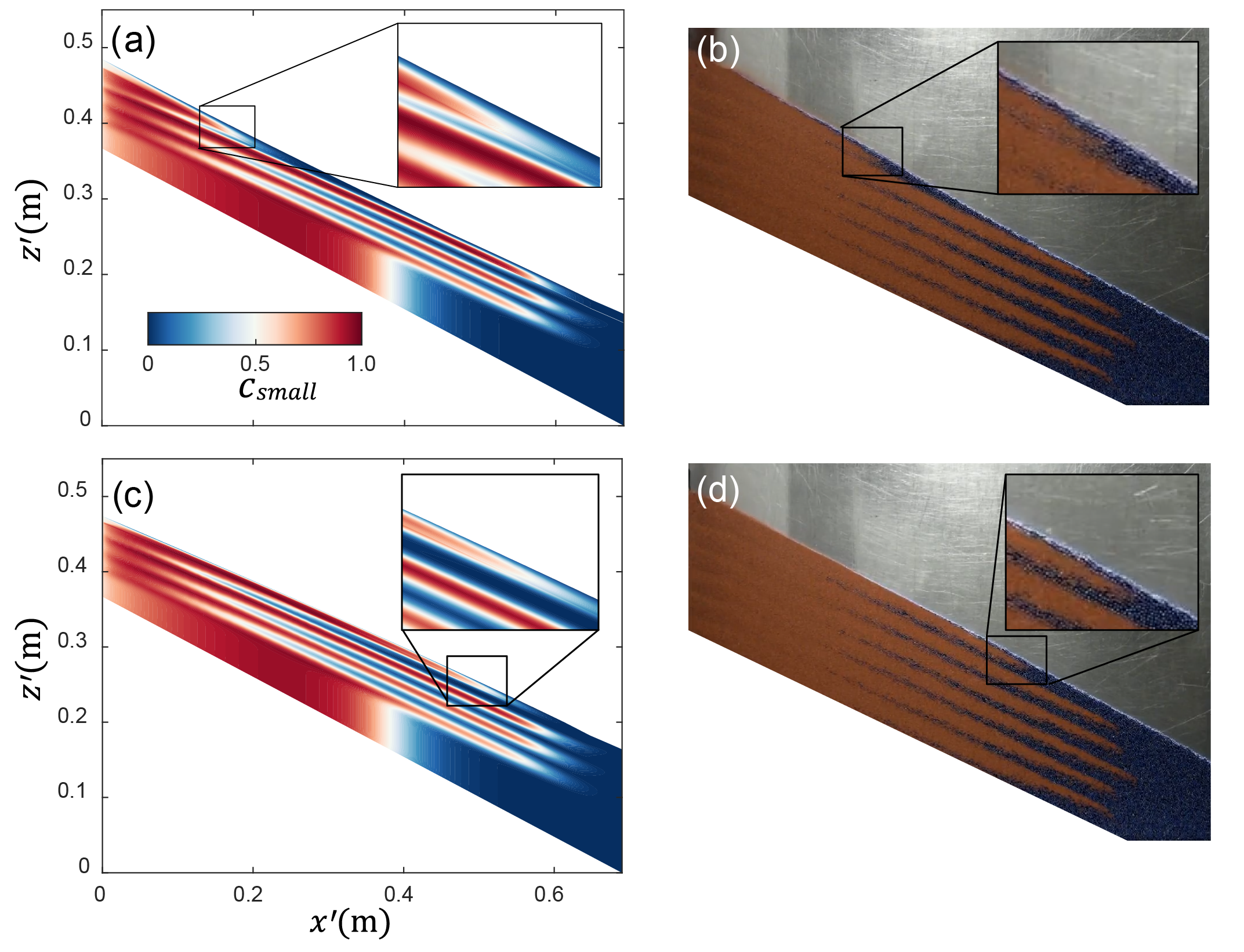}
	\caption[Model validation]{Comparison of continuum model predictions (left column) to experimental results~\cite{Xiao2017_2}  (right column) in a $W=69$\,cm wide and $T=1.27$\,cm thick heap under modulated flow conditions for an equal volume mixture of 2\,mm blue and 0.5\,mm red glass spheres at (a, b) 3\,s into the slow-to-fast transition and at (c, d) 10\,s into the fast-to-slow transition.  Color bar in (a) indicates the small-particle concentration.  Modulation parameters: $q_\mathrm{s} = 2 $\,cm$^2$/s, $t_\mathrm{s} = 20$\,s, $q_\mathrm{f} = 23.6 $\,cm$^2$/s, and $t_\mathrm{f} = 5$\,s. Note that the continuum model (left column) starts with a steady feed rate (lower portion of the flow domain) followed by slightly more than three iterations of feed rate modulation, while the experiments include six iterations of feed rate modulation without an initial steady feed rate period.}
	\label{fig:validation}
\end{figure*}

Before exploring the dynamics of modulated-flow-driven stratification using the continuum model, we first validate the model against experiment. Specifically, model predictions are compared with a previous experiment~\cite{Xiao2017} at the same operating conditions and at two different times during bed formation in Fig.~\ref{fig:validation}. For the model results, red corresponds to high small-particle concentration, blue to high large-particle concentration, and white indicates an equal volume mixture of the two species. The continuum model (left column) starts with a steady feed rate period, evident as simple segregation with small red particles upstream and large blue particles downstream in the lower portion of the computational domain (where particles are first deposited), before slightly more than three iterations of feed rate modulation occur, evident as the sloped stratification pattern in the upper portion of the domain. The experiment includes six iterations of the feed rate modulation but does not start with a steady feed rate interval. 

The stratified segregation pattern generated by the continuum model (upper portion of the domain in the left column) matches the pattern in the experiment (right column) in terms of the extent each segregated layer penetrates into the upstream and downstream regions, the inclination of the stratified layers relative to the free surface angle, and the change in layer thickness vs.\ streamwise position on the heap (i.e.,\ large blue particle layers increase in thickness downstream, while small red particle layers maintain a nearly constant thickness). 

Not only does the stratification pattern of the particles deposited on the static heap match between the continuum model and the experiment, but the details of how the pattern forms during slow-to-fast and fast-to-slow feed rate transitions also match. When the fast feed phase starts, a slight ``positive wedge" forms at the top of the heap and propagates toward the downstream endwall. As the material flows downstream, segregation in the wedge causes an excess of large blue particles to accumulate at the leading edge of the layer of small (red) particles, as shown in the magnified regions in Fig.~\ref{fig:validation}(a, b) $3$\,s into the fast feed phase (this concentration of large particles at the leading edge of the small particles is reminiscent of similar behavior observed in avalanches~\cite{Gray1997}). These large particles are followed and buried by small particles in the flowing layer, resulting in deposition of large particles onto the upstream portion of the heap. It is important to note that for these conditions the wedge front is already almost three-quarters of the way down the slope based on Eq.~\ref{front_position_eqn}. Thus, the leading edge of the small particles trails behind the wedge front. 

Similarly, when the slow feed phase begins, the height of the free surface near the feed zone decreases slightly, forming a slight ``negative" wedge in the upstream region of the heap. As the negative wedge propagates toward the downstream endwall, previously deposited small particles in the upstream regions of the bed are eroded into the flowing layer, flow downstream, and then deposit back onto the static bed further downstream than where they were initially eroded. This process is visible in the magnified regions of Fig.~\ref{fig:validation}(c, d) $10$\,s into the slow feed phase where the leading edge of the layer of small (red) particles has advanced more than two-thirds of the way down the slope. 

Close agreement between the overall pattern and details of the pattern forming process in the continuum model predictions and experiments is also observed for other combinations of the modulations parameters (i.e., $q_j$ and $t_j$).  Having confirmed that the model can accurately reproduce experiments, we now use it to better understand the stratification process and explore the pattern dependence on the feed-rate-modulation parameters. 

\section{Transient segregation during single feed rate transitions}

\begin{figure}[t]
	\centering
	\includegraphics[width=0.95\columnwidth]{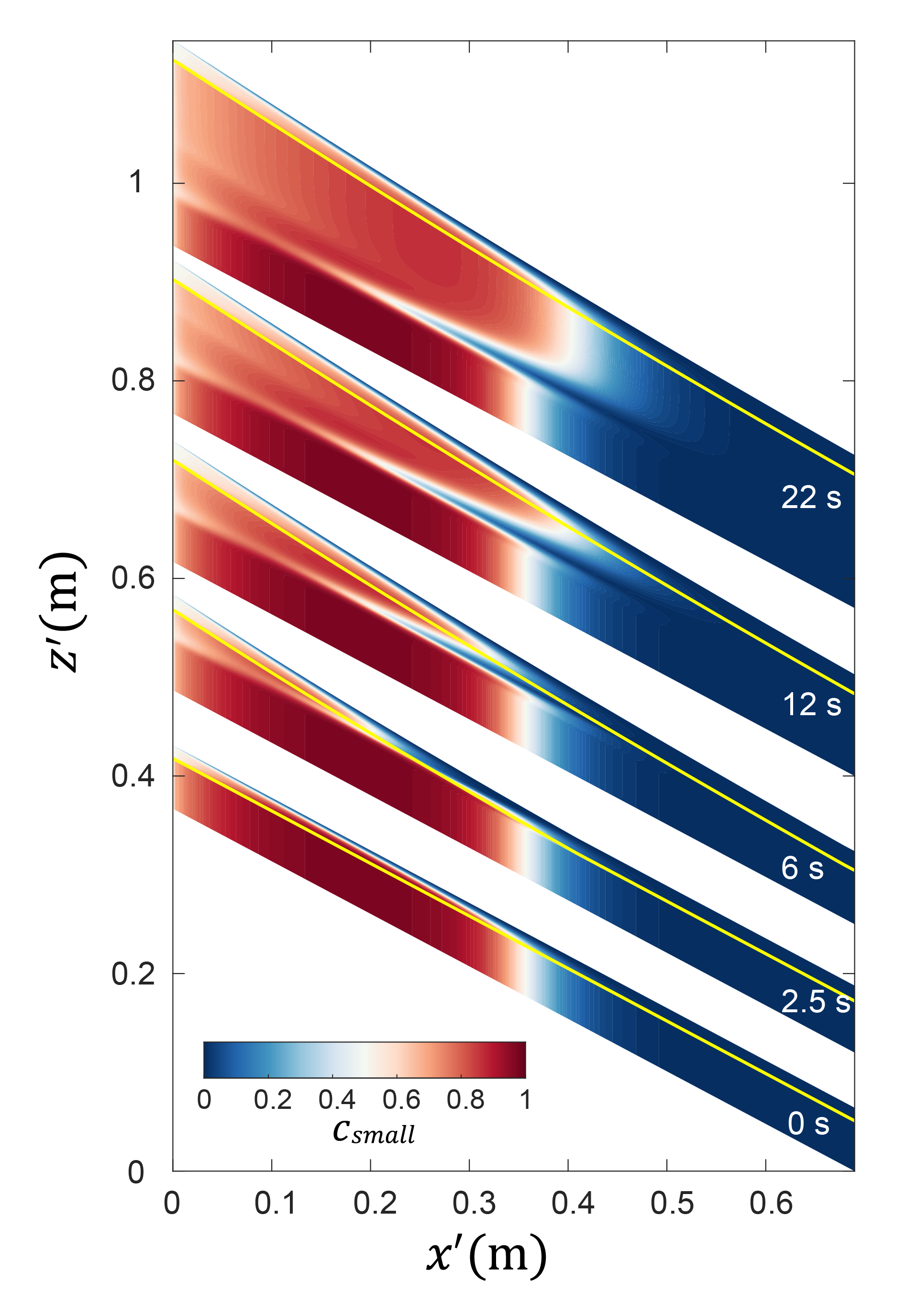}
	\caption[Layer formation in continuum model after switch from slow-to-fast feed rate]{Layer formation due to wedge propagation after transition from a slow ($q_\mathrm{s} = 12 $\,cm$^2$/s) to a fast ($q_\mathrm{f} = 36 $\,cm$^2$/s) feed rate at $t=0$ for different times $t$ (increasing from bottom to top) according to the continuum model for an equal volume mixture of 2\,mm blue and 0.5\,mm red glass spheres with $W=69$\,cm. Small-particle concentration, $c_\mathrm{small}$, is indicated by color bar, yellow curves indicate the bottom of the flowing layer, $z^\prime = z_\mathrm{b}^\prime$. Times correspond to $t/\tau=0, 0.44, 1.05, 2.11, 3.86$.}
	\label{Slow_to_fast}
\end{figure}

Before applying the model to the more complicated case of the periodically modulated feed rate, we first consider the simpler case of a single feed rate transition in order to better understand the propagation of the change in flow rate down the heap and its effect on segregation. Figure~\ref{Slow_to_fast} shows the segregation in terms of the concentration of small particles, $c_\mathrm{small}$, during downslope wedge propagation for a slow to a fast feed rate transition at five different times for an equal volume mixture of small and large particles.  The yellow curve in each concentration field indicates the instantaneous interface at the bottom of the flowing layer, $z^\prime = z_\mathrm{b}^\prime$, at the current time step of the computations.  Particles deposited on the heap lie below this interface; particles flowing down the heap are above it. 

In steady state (i.e., $t=0$ in Fig.~\ref{Slow_to_fast}), the concentration in the static portion of the heap (beneath the flowing layer) varies only in the streamwise direction.  In the upstream portion, mostly small particles are deposited (large $c_\mathrm{small}$, red) due to rapid segregation in the flowing layer, although $c_\mathrm{small}$ is slightly less at the extreme upstream end of the heap where there is too little time for some large particles to segregate upward before depositing on the static heap. Approximately half-way down the slope (for this equal volume mixture), the flowing layer becomes depleted of small particles, and $c_\mathrm{small}$ diminishes rapidly so that the flowing layer is composed entirely of large particles (blue), which are then deposited exclusively on the remainder of the static heap. The resulting streamwise-segregated pattern generated by the model is typical of heaps at moderate feed rates~\cite{Fan2012}. 

Now consider how a sudden increase in the feed rate alters the concentration field.  Shortly after switching to the fast feed rate ($t>0$), a positive wedge of material forms at the upstream end of the flowing layer, evident at $t = 2.5$\,s in Fig.~\ref{Slow_to_fast}, which corresponds to $t/\tau=0.44$, where $\tau=W^2/C=5.7$\,s. At this time, the wedge front is at $x^{\prime}_\mathrm{wf}=0.46$\,m ($x^{\prime}_\mathrm{wf}/W=0.67$) according to Eq.~\ref{front_position_eqn}. However, the leading edge of the small particles is well behind the wedge front at $x^{\prime}\approx 0.2$. This is because as the wedge front advances downstream, large particles segregate upward toward the free surface of the wedge, while small particles percolate downward toward the bottom of the flowing layer in the usual manner. Since the streamwise velocity of large particles on the free surface of the wedge is greater than that for small particles deeper in the flowing layer and greater than the wedge front velocity, large particles are conveyed to the forward portion of the wedge. (Similar conveying behavior occurs in segregating avalanche flows~\cite{Gray2009, Xiao2017}.) The large particle concentration in the forward portion of the wedge grows until the flowing layer there consists almost entirely of large particles. Consequently, large particles deposit from the wedge front onto the static portion of the heap where only small particles had previously deposited during the preceding steady slow feed rate ($t\le 0$). These large particles are then covered by small particles as the front moves further downstream ($t=6$\,s and $t=12$\,s), which forms a large particle enriched layer in the upstream portion of the heap visible for $t\ge 6$\,s in the figure. 

After the wedge front reaches the downstream boundary of the heap, corresponding to $t=\tau=5.7$\,s, the flowing layer relaxes back to steady-state (here for $t\gtrsim 12$\,s), but at the higher feed rate $q_\mathrm{f}$. In steady state at $q_\mathrm{f}$, the transition between small and large particles (i.e., the white region corresponding to\ $c_\mathrm{small}=0.5$) is slightly broader and located further downstream, and the small particle concentration in the upstream region is reduced compared to the slow feed rate, see Fig.~\ref{Slow_to_fast} at $t = 22$\,s.  These three differences are due to the increased importance of advection and diffusion relative to segregation at the higher feed rate~\cite{Fan2014a}.

The most significant result shown in Fig.~\ref{Slow_to_fast} at $t = 22$\,s is that after a single change in the feed rate from slow-to-fast there is a single layer of large particles extending significantly further upstream on the heap than the steady-state position of the boundary between small and large particle dominated regions for either the slow or fast feed rate. The full upstream extent of this layer to $x^{\prime}\approx 0.2$ is fixed by $t = 6$\,s, which corresponds to $t/\tau=1.05$, the approximate time for the wedge front to reach the downstream wall. The thickness of the large particle layer grows for some time after this as the flow relaxes toward its steady state condition along the entire length of the surface after the wedge front reaches the downstream wall, but its final thickness is established by $t = 12$\,s ($t/\tau=2.11$), as is evident by comparing it to the situation at $t = 22$\,s. Thus, it appears that the upstream extent of the large particle layer is established by the time the wedge front reaches the downstream wall at $\tau$ but continues to thicken until about $2\tau$.

\begin{figure}[t]
	\centering
	\includegraphics[width=0.95\linewidth]{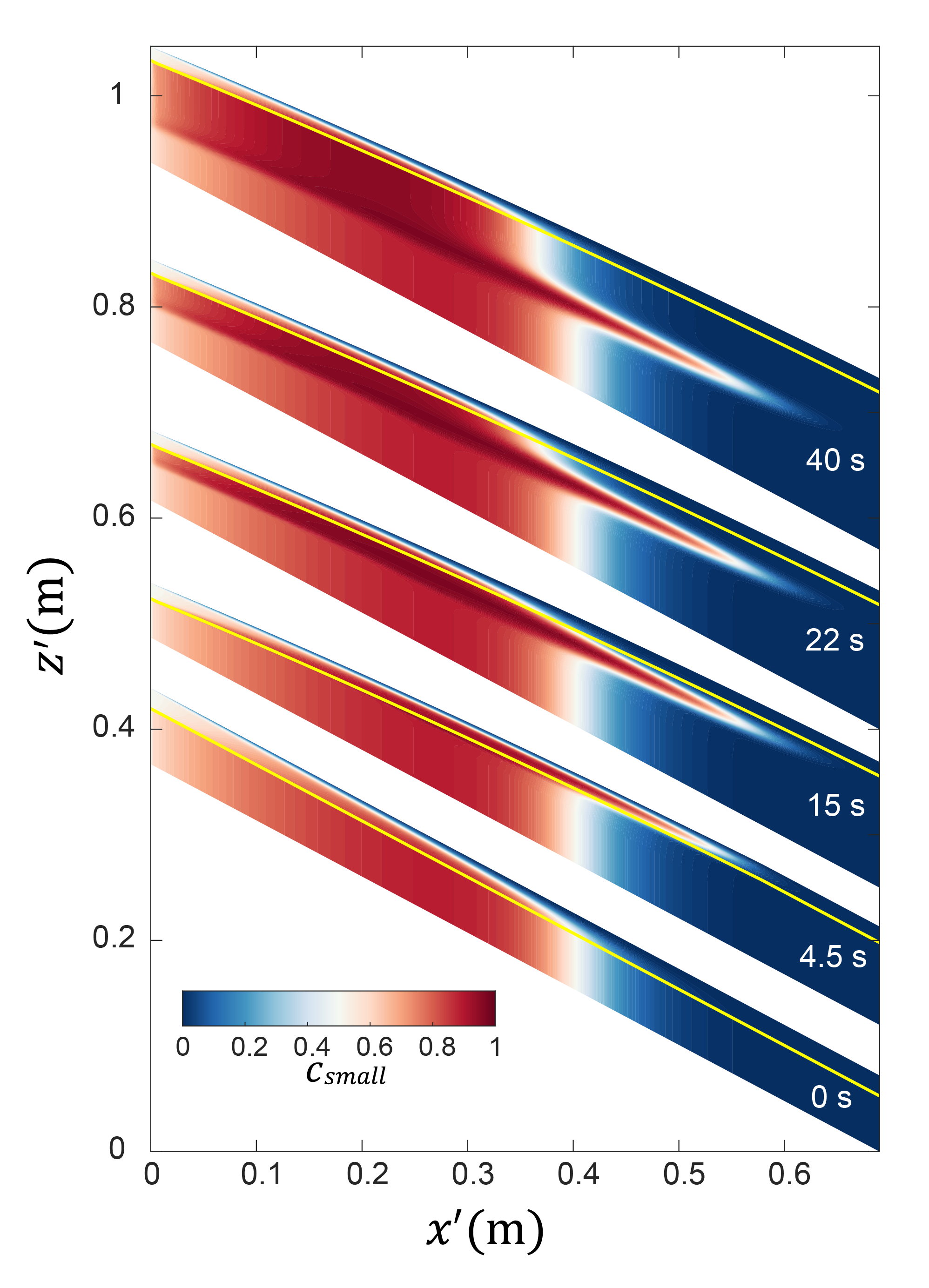}
	\caption[Layer formation after transition from fast flow to slow flow rate]{Layer formation due to wedge propagation after transition from a fast ($q_\mathrm{f} = 36 $\,cm$^2$/s) to a slow ($q_\mathrm{s} = 12 $\,cm$^2$/s) feed rate at $t=0$ for different times $t$ (increasing from bottom to top) according to the continuum model for an equal volume mixture of 2\,mm blue and 0.5\,mm red glass spheres with $W=69$\,cm. Small-particle concentration, $c_\mathrm{small}$, is indicated by color bar, yellow curves indicate the bottom of the flowing layer, $z^\prime = z_\mathrm{b}^\prime$. Times correspond to $t/\tau=0, 0.79, 2.63, 3.86, 7.02$.}
	\label{fast_to_slow}
\end{figure}

Figure~\ref{fast_to_slow} shows the analogous layer-formation process for a fast-to-slow feed rate transition. The initial fully segregated condition for the static bed at $t=0,$ is similar to that in Fig.~\ref{Slow_to_fast}, except that the transition from small (red) to large (blue) particles is a bit wider and further downstream due to the faster feed rate. After switching to the slow feed rate at $t=0,$ the surface height in the upstream portion of the flowing layer decreases slightly (evident at $t=4.5$\,s and $t=15$\,s) due to the associated surface relaxation, forming a negative wedge. As a result, previously deposited small-particle-enriched material re-enters the flowing layer via erosion of the static bed (darker red in the lower part of the upstream portion of the flowing layer), and segregation in the flowing layer is amplified due to the higher concentration of small-particles and the slower flow rate compared to when they were first deposited.  As the negative wedge advances downstream and the heap adjusts to the lower repose angle, the flowing layer continues to be enriched by small particles via bed erosion.  This allows the flowing layer to deposit small particles further downstream before being depleted of small particles, as is evident in the figure for $t=4.5$\,s and  $t=15$\,s.  The ultimate downstream extent of the small particle layer at $x^{\prime}\approx 0.6$\,m is almost established by $t=4.5$\,s, which corresponds to $t/\tau=0.79$. But like the upstream large particle layer for the slow-to-fast feed rate transition, the small particle layer grows thicker for $t>\tau$, evident at $t=15$\,s ($t/\tau=2.63$), as the flow relaxes toward its steady state condition along the entire length of the surface after the negative wedge front reaches the downstream wall. After the formation of the small particle layer, steady-state heap growth resumes but at the slower flow rate and shallower slope, as shown at $t = 40$\,s in the figure.  Accordingly, the streamwise concentration transition at $c_\mathrm{small}=0.5$ is located further upstream and is narrower than for $q_\mathrm{f}$.  

The net result of the fast-to-slow feed rate transition, evident in Fig.~\ref{fast_to_slow} at $t=40\,s$, is a layer of deposited small particles that extends significantly further downstream than the steady-state position of the boundary between small and large particle dominated regions at either the slow or fast feed rate.

\section{Stratified segregation patterns: parameter dependence}

As illustrated in Sec.~\ref{sec_validation}, when the two stratification processes described in Figs.~\ref{Slow_to_fast} and \ref{fast_to_slow} occur sequentially and repeatedly, large-particle layers deposit further upstream during a slow-to-fast transition (Fig.~\ref{Slow_to_fast}) and small-particle layers deposit further downstream during a fast-to-slow transition (Fig.~\ref{fast_to_slow}), resulting in a periodic layering of the large and small particle species.  The continuum model developed in Secs.~\ref{sec:segmodel} and ~\ref{sec:trans} is useful in exploring the parameter dependence of the stratified segregation patterns.  

\begin{figure*}[t]
	\centering
	\includegraphics[width=\textwidth]{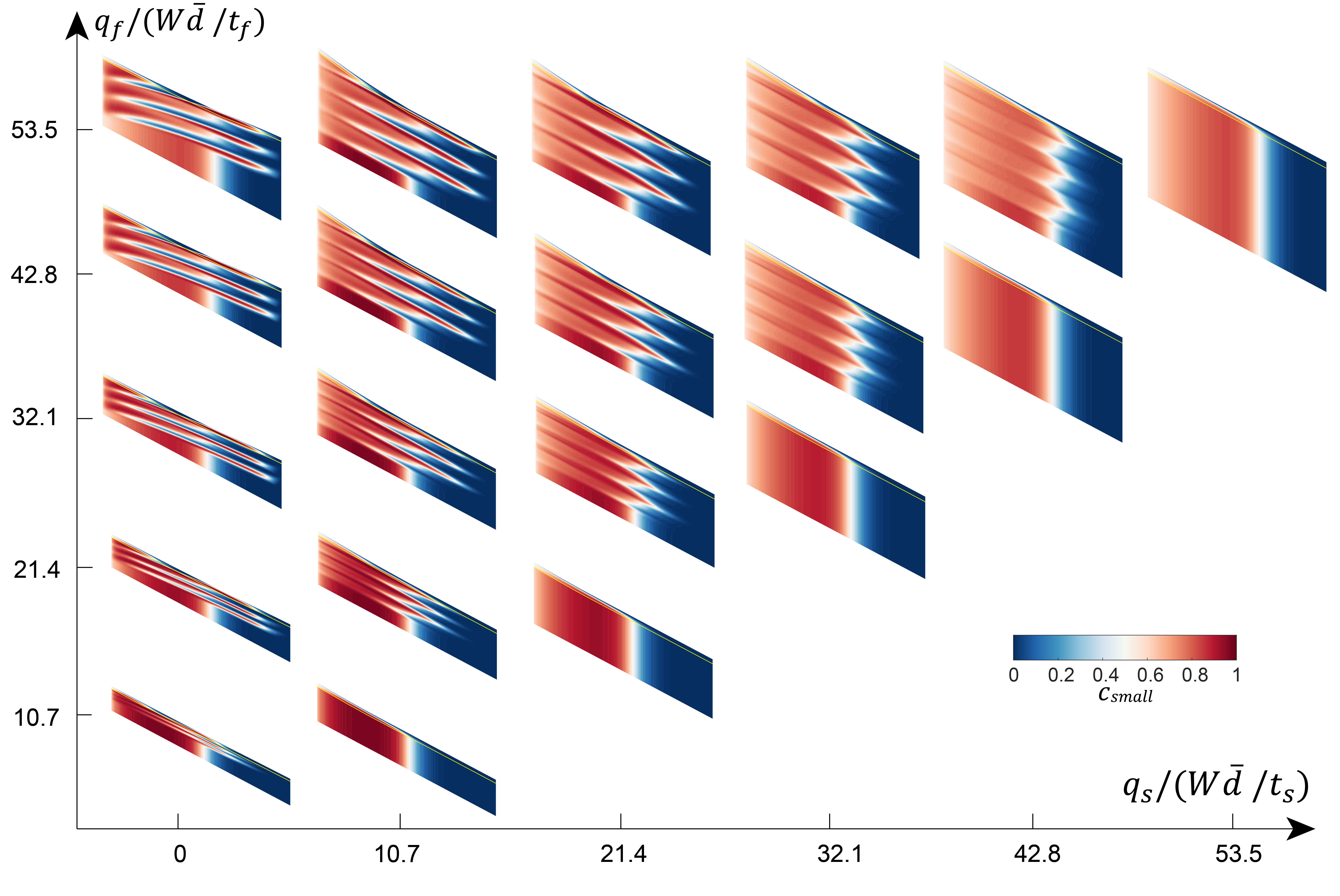}
	\caption{Effect of varying fast and slow feed rates on stratification for $q_\mathrm{s}\in \{0, 12, 24, 36, 48, 60\}$\,cm$^2$/s and $q_\mathrm{f}\in \{12, 24, 36, 48, 60\}$\,cm$^2$/s with equal feed durations ($t_\mathrm{f}=t_\mathrm{s}=1.41\tau=8$\,s) after three full cycles of modulation ($48$\,s) following an initial period of steady state heap growth to a target bed height as described in Sec.~\ref{sec:trans}. The initial steady state is formed at a feed rate of $q_\mathrm{s}$ for all cases except for when $q_\mathrm{s}= 0$, in which case the initial feed rate is $q_\mathrm{f}$. }
	\label{varying_q}
\end{figure*}

Figure~\ref{varying_q} shows segregation patterns resulting from repeated slow-to-fast and fast-to-slow transitions for twenty combinations of $q_\mathrm{f}$ and $q_\mathrm{s}$ with $t_\mathrm{f}=t_\mathrm{s}=8$\,s, which corresponds to $t_j/\tau=1.41$. The two-dimensional feed rate, $q_j$, is nondimensionalized with the horizontal width of the heap, $W$, average particle diameter, $\bar{d}$, and feed rate duration, $t_j$, such that $\tilde{q}_j = q_j/(W \bar{d}/t_j)$ . This can be thought of as the two-dimensional area of material added to the heap during each feed rate interval, $q_j t_j$, measured in units of the area of a single layer of particles, $W \bar{d}$. In each case, the system starts from a base deposited in steady flow at the slow feed rate $q_\mathrm{s}$ (except when $q_\mathrm{s}=0$ for which the steady initial flow is at $q_\mathrm{f}$) and then  switches repeatedly between fast and slow feed rates every 8\,s.  Only feed rate combinations with $\tilde{q}_\mathrm{f} \ge \tilde{q}_\mathrm{s}$ are shown, since the patterns developed for $\tilde{q}_\mathrm{f} < \tilde{q}_\mathrm{s}$ are identical after the second feed rate switch.

When $\tilde{q}_\mathrm{f}= \tilde{q}_\mathrm{s}$, the feed rates are equal so no layering occurs.  As discussed in the previous section, with increasing feed rate (moving up the diagonal), the equal concentration boundary (white) moves slightly downstream, the streamwise extent of the concentration transition from small to large particle dominated regions increases, and the degree of segregation decreases in the upstream portion of the heap as the relative influences of advection and diffusion in the flowing layer increase relative to that of segregation.

Consider now the limiting case where one of the feed rates is zero, here $\tilde{q}_\mathrm{s}=0$, shown in the first column in Fig.~\ref{varying_q}. The slow-fast transition (in this case, zero-fast transition) results in a layer of large particles extending relatively far up the slope, while the fast-slow transition (fast-zero transition) results in a layer of small particles extending down the slope. Compared to the other cases in Fig.~\ref{varying_q} where both $\tilde{q}_\mathrm{s}$ and $\tilde{q}_\mathrm{f}$ are non-zero, the length of the stratified layers is longest when one of the feed rates is zero.  This is consistent with the observation in experiments that the length of the layer increases with decreasing $\tilde{q}_\mathrm{s}$~\cite{Xiao2017}. Furthermore, this result is similar to that for stratification in heaps at low constant feed rates where the flow down the slope is quasi-periodic (on-off)~\cite{Makse1997, Williams1976, Gray1997, Baxter1998, Fan2012}. We note, however, that when the duration $t_\mathrm{f}$ of the non-zero feed rate ($q_\mathrm{f}$) is too short, corresponding to small values of $\tilde{q}_\mathrm{f}$, the amount of material deposited is insufficient to form a long layer of material (e.g., $\tilde{q}_\mathrm{f}=10.7$ or 21.4, $\tilde{q}_\mathrm{s}=0$). Also note that the angle of the stratified layer with respect to the angle of repose differs slightly for $\tilde{q}_\mathrm{s}=0$ compared to the other cases due to the different initial condition (steady feed at $\tilde{q}_\mathrm{f}$ instead of $\tilde{q}_\mathrm{s}$).

When the two feed rates are non-zero and differ from one another (i.e.,\ $\tilde{q}_\mathrm{f}>\tilde{q}_\mathrm{s}$), layering still occurs. If $\tilde{q}_\mathrm{f}\sim \tilde{q}_\mathrm{s}$ (e.g., $\tilde{q}_\mathrm{f}=53.5, \tilde{q}_\mathrm{s}=42.8)$, the streamwise extent of layering is small, but if $\tilde{q}_\mathrm{f} \gg \tilde{q}_\mathrm{s}$, (e.g., $\tilde{q}_\mathrm{f}=53.5$, $\tilde{q}_\mathrm{s}=10.7$), the streamwise extent of layering is nearly as large as when $\tilde{q}_\mathrm{s}=0$.  Large relative differences between $\tilde{q}_\mathrm{f}$ and $\tilde{q}_\mathrm{s}$ are needed for strong layering to both extend the small particle layer further downstream due to increased erosion and flowing layer segregation during the fast-to-slow feed transition and to bring the large particle layer further upstream by increasing the deposition of large particles during the slow-to-fast transition.  However, if too much material is deposited during a feed interval, the flow reaches steady state with the small particles depositing upstream and large particles depositing downstream according to the steady-state pattern. This increases the small particle layer thickness in the upstream region and the large particle layer thickness in the downstream region without increasing the thicknesses of the penetrating layers (upstream or downstream) or the penetration of those layers (upstream or downstream).  This effect is visible in Fig.~\ref{varying_q} for the largest fast feed rate ($\tilde{q}_\mathrm{f}=53.5$, top row) compared to the row just below it, particularly for $\tilde{q}_\mathrm{s}=0,$ 10.7, and 21.4. In these cases, the penetration of the layers upstream and downstream does not change from one row to the other and the vertically invariant concentration regions matching the steady flow (diagonal) conditions are visible after every slow-to-fast feed transition.  In addition, the total amount of material deposited per modulation cycle is $\sum q_j t_j,$ as is evident in the increasing thickness of each layer of the heap moving up (higher $\tilde{q}_\mathrm{f}$) or to the right (higher $\tilde{q}_\mathrm{s}$) in the figure.  Thus, the thinnest and deepest penetrating layers occur on the left side of the figure where the ratio of the fast-to-slow feed rate is large but the average feed rate is low. The optimal layering is a balance between penetration (large enough relative difference between $\tilde{q}_\mathrm{f}$ and $\tilde{q}_\mathrm{s}$) and avoiding reaching steady state ($\tilde{q}_\mathrm{f}$ or $\tilde{q}_\mathrm{s}$ is too large). When $\tilde{q}_\mathrm{f}$ and $\tilde{q}_\mathrm{s}$ are both small, the penetration is reduced (e.g., $\tilde{q}_\mathrm{f}=21.4, \tilde{q}_\mathrm{s}=10.7$), but when $\tilde{q}_\mathrm{f}$ or $\tilde{q}_\mathrm{s}$ is too large, the layers thicken (e.g., $\tilde{q}_\mathrm{f}=53.5, \tilde{q}_\mathrm{s}=10.7$) but not for the penetrating portions of the layers.

\begin{figure*}[t]
	\centering
	\includegraphics[width=\textwidth]{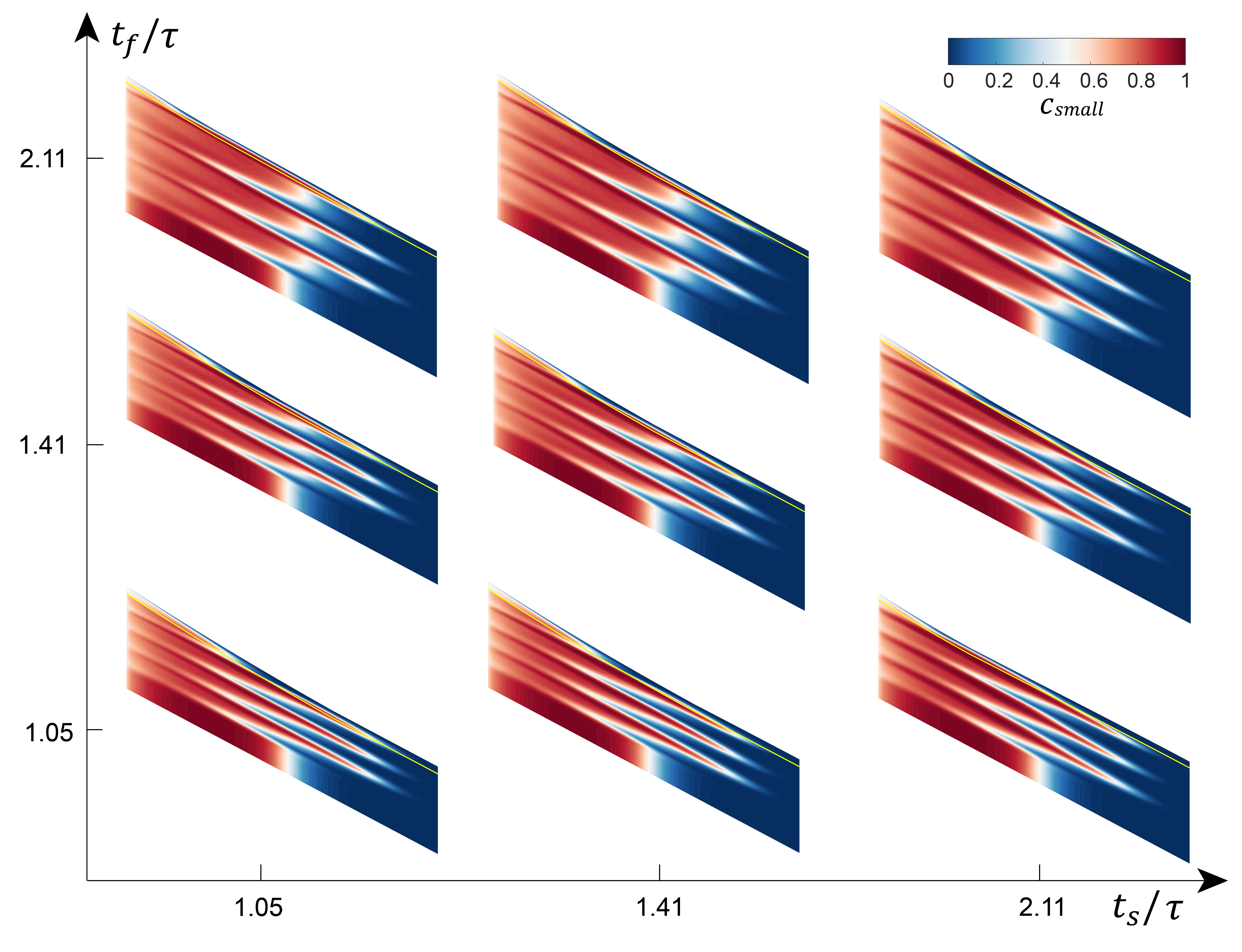}
	\caption{Effect of feed rate duration on stratification starting from a steady slow feed rate $q_\mathrm{s} = 12 $\,cm$^2$/s and then alternating between $q_\mathrm{f}= 36 $\,cm$^2$/s and $q_\mathrm{s} = 12 $\,cm$^2$/s for non-dimensionalized feed durations $t_j/\tau \in \{1.05, 1.40, 2.11\}$.}
	\label{Varying cycle times}
\end{figure*}

In Fig.~\ref{varying_q} the two feed rate durations are constant and equal, $t_\mathrm{f}=t_\mathrm{s}=1.41\tau=8$\,s.  In Fig.~\ref{Varying cycle times} both the fast and slow durations ($t_\mathrm{f}$ and $t_\mathrm{s}$) are varied while the feed rates are fixed at $q_\mathrm{s}=12$\,cm$^2$/s and $q_\mathrm{f}=36$\,cm$^2$/s, corresponding to $(\tilde{q}_\mathrm{s}=10.7, \tilde{q}_\mathrm{f}=32.1)$ in Fig.~\ref{varying_q}.  Since our computational approach only tracks the position of a single wedge front at a time, we only consider feed durations that exceed the time for the wedge front to reach the downstream wall, $\tau;$ that is, $t_j/\tau \geq 1.$ Feed durations shorter than this require computations in which a second wedge front starts to propagate before the previous wedge front reaches the downstream wall. While this is computationally feasible, we do not pursue this case here. 
%other than to comment that situations with $t_j/\tau < 1$ would result in incomplete formation of the layers so they would be shorter in their extent upstream or downstream. I'm not sure this is correct as the earlier wedge front would not ``know" about the change in feed rate for some time.

For all simulated combinations of $t_\mathrm{f}$ and $t_\mathrm{s}$ in Fig.~\ref{Varying cycle times}, relatively strong layering is observed due to the relatively large difference in feed rates, $\tilde{q}_\mathrm{f} = 3 \tilde{q}_\mathrm{s}.$  Very little, if any, change in the upstream extent of the large particle layer associated with varying the fast feed rate interval is observed because $t_\mathrm{f}/\tau > 1.$ The large particle layer in the upstream portion of the bed is always fully deposited before the end of the feed interval, and its deposition is thus unaffected by the duration of $t_\mathrm{f}$.  Neither is the upstream extent of the large-particle enriched layer affected by $t_\mathrm{s},$ because in all cases erosion associated with the negative wedge is never deep enough to reach the previously deposited large particles.  The downstream extent of small-particle enriched layer is similarly insensitive to $t_\mathrm{s},$ again because the erosive phase of the fast-to-slow transition is completed before the slow feed rate phase ends. 

The layer thickness increases with increasing $t_\mathrm{f}$ or $t_\mathrm{s},$ but more rapidly for $t_\mathrm{f}$ due to its larger associated feed rate, although the upstream penetration of the large particle layer and downstream penetration of the small particle layer remain relatively unchanged in both streamwise extent and thickness. This is most easily seen in the left column for $t_\mathrm{s}/\tau =1.05$. As $t_f/\tau$ increases, the additional time, $t_\mathrm{f}>\tau$, following the slow to fast feed rate transition that forms the upstream penetrating large particle layer results in steady-state deposition of particles, consistent with Fig.~\ref{Slow_to_fast}. Similarly, for the bottom row, increasing $t_\mathrm{s}/\tau$ results in steady-state deposition of particles following the fast to slow feed rate transition that forms the downstream penetrating small particle layer, consistent with Fig.~\ref{fast_to_slow}. This is more difficult to see than the case for the slow to fast transition because fewer particles are deposited at the slower feed rate in time $t_\mathrm{s}$; it is barely visible as the slightly increasing curvature of the interface between large and small particles at the top of the small particle layer as $t_\mathrm{s}/\tau$ increases much like that evident in Fig.~\ref{fast_to_slow} for the upper surface of the small particle layer at $t=22$\,s. However, in both cases the penetrating portion of the layer remains similar in both thickness and extent, regardless of $t_j/\tau$. Although increasing $t_j/\tau$ increases the overall layer thickness, the regions of large particles penetrating upstream and small particles penetrating downstream are quite similar for all $t_j/\tau$; it is only the region of steady-state deposition following the transition that changes the overall layer thickness. 

Of course, uniform mixtures of segregating particles are usually desirable in industrial processes.  Modulating the feed rate to generate layers of small and large particles can enhance the overall mixture uniformity on heaps even when materials are strongly segregating (e.g., due to large size or density ratios) or when sufficiently high feed rates to avoid segregation  are impractical. However, the layering resulting from intentional modulation of the feed rate is imperfect -- there is still a general tendency for small particles to deposit on the upstream portion of the heap and large particles to deposit on the downstream portion. In addition, the layers' streamwise extent and thickness depend on the modulation parameters.  As we have shown in this section, the length of the layer increases as the ratio of $\tilde{q}_\mathrm{s}$ to $\tilde{q}_\mathrm{f}$ decreases, and layer thickness decreases with decreases in the total material deposited, $\sum q_j t_j$, during either the slow or fast phase of the feed modulation. 

\begin{figure}[t]
	\centering
	\includegraphics[width=\columnwidth]{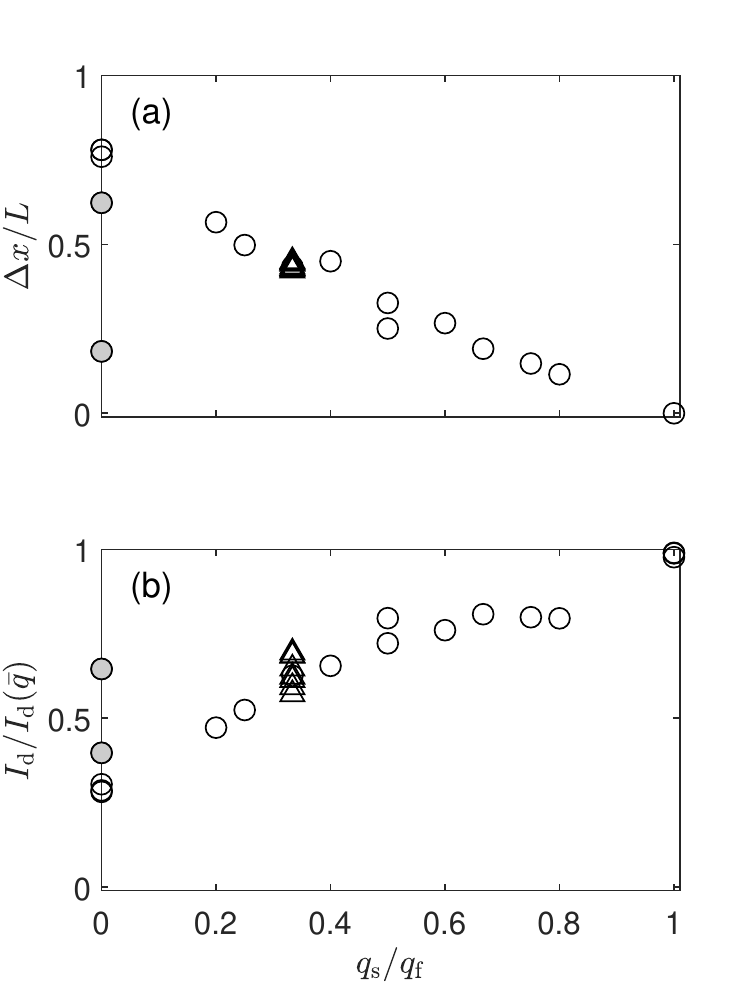}
	\caption{Effect of feed rate modulation on (a) layer penetration depth, $\Delta x/L$, and (b) normalized Danckwerts' intensity of segregation~\cite{Danckwerts1952}, $I_\mathrm{d}/I_\mathrm{d}(\bar{q})$, (see text) for all cases included in Figs.~\ref{varying_q} ($\circ$)  and~\ref{Varying cycle times} ($\triangle$). The two shaded circles indicate conditions with $q_\mathrm{f} \in \{12, 24\}$ where insufficient material is deposited to form a full layer (see text).}
	\label{fig:penetration}
\end{figure}

To quantify the dependence of the layers on the feed modulation parameters, the penetration length of the segregated layer $\Delta x$ is quantified in terms of the streamwise extent of the $c_\mathrm{small}=0.5$ (white) contour for the cases in Figs.~\ref{varying_q} and~\ref{Varying cycle times}. The dependence of the normalized length of the segregated layer, $\Delta x/L$, for these cases is shown in Fig.~\ref{fig:penetration}(a) as a function of $q_\mathrm{s}/q_\mathrm{f}$. Consistent with experiments~\cite{Xiao2017}, the greatest layer penetration is achieved with on-off modulation of the feed rate (i.e. ${q}_\mathrm{s}=0$). The length of the stratification layer decreases linearly from $\Delta x/L\approx 0.8$ at $q_\mathrm{s}/q_\mathrm{f}=0$ to $\Delta x/L=0$ at $q_\mathrm{s}/q_\mathrm{f}=1$, apart from the two shaded data points corresponding to situations where $\tilde{q}_\mathrm{f}$ is too small to fully form a stratification layer ($\tilde{q}_\mathrm{f}=10.7$ and 21.4 in Fig.~\ref{varying_q}). The cluster of data points at $q_\mathrm{s}/q_\mathrm{f}=0.33$ corresponds to the data in Fig.~\ref{Varying cycle times}, demonstrating that the duration $t_j$ does not alter the length of the stratification layer, provided that $t_j>\tau$.

It is also possible to consider the degree of segregation in terms of the feed modulation parameters. Here we characterize a region representative of the material deposited on the heap during modulated flow consisting of two deposited stratification layers just below the flowing layer in terms of the Danckwerts segregation parameter~\cite{Danckwerts1952}, calculated as:
\begin{equation}
I_\mathrm{d} = \frac{1}{\bar{c}(1-\bar{c})L} \int_{0}^{L}  \frac{l(x)(c(x)-\bar{c})^2}{\bar{l}} dx,
\label{Danckwerts}
\end{equation}
where $c(x)$ is the average local species concentration in a slice of length $l(x)$ that is normal to the free surface of the heap and extends from the center of the first layer to the center of the third layer at streamwise position $x$, $\bar{c}=0.5$ is the average concentration of one of the particle species, and $\bar{l}$ is the average slice length. The slice length equals twice the layer spacing except at the upstream and downstream ends of the heap where it is reduced because of the vertical end walls. The segregation index, $I_\mathrm{d}$, varies from 0 for perfectly mixed to 1 for completely segregated. We normalize the measured segregation index by the segregation index for the unmodulated case with the same average feed rate, $I_\mathrm{d}(\bar{q})$, where $\bar{q}= \sum q_j t_j/\sum t_j$ is the average feed rate for the modulated case. This requires fitting a curve to the segregation index data for the unmodulated flow cases on the diagonal in Fig.~\ref{varying_q} (i.e.\ $I_\mathrm{d}(\bar{q})= I_0 e^{\bar{q}/q_0}$ with $I_0=0.85$ and $q_0=88$), for which $I_\mathrm{d}$ varies from $0.76$ for the lowest flow rate to $0.44$ for the highest flow rate, and using $I_\mathrm{d}(\bar{q})$ for the value of $\bar{q}$ corresponding to $q_\mathrm{s}/q_\mathrm{f}$ for the other modulated flow cases in Figs.~\ref{varying_q} and~\ref{Varying cycle times}. This normalization isolates the effect of the modulation alone over the steady feed rate case by accounting for the impact of the average feed rate, $\bar{q}$, on $I_\mathrm{d}$.

The normalized segregation index measured for the cases in Figs.~\ref{varying_q} and~\ref{Varying cycle times} are shown in Fig.~\ref{fig:penetration}(b) as a function of $q_\mathrm{s}/q_\mathrm{f}$. The greatest overall mixing (smallest $I_\mathrm{d}$) occurs with on-off modulation of the feed rate (i.e.\ ${q}_\mathrm{s}=0$), and the particles become more segregated with increasing $q_\mathrm{s}/q_\mathrm{f}$ to a maximum of $I_\mathrm{d}/I_\mathrm{d}(\bar{q})= 1$ at $q_\mathrm{s}/q_\mathrm{f}=1$. Again, the two shaded data points correspond to situations where $\tilde{q}_\mathrm{f}$ is too small to fully form a layer ($\tilde{q}_\mathrm{f}=10.7$ and $21.4$ in Fig.~\ref{varying_q}), so they are more segregated (larger $I_\mathrm{d}$) than other cases at $q_\mathrm{s}/q_\mathrm{f}=0$. The remaining cases for $q_\mathrm{s}/q_\mathrm{f}=0$ are not perfectly mixed because the layers do not extend the full length of the heap, see Fig.~\ref{fig:penetration}(a). There is some spread in the segregation index for the cluster of data points at $q_\mathrm{s}/q_\mathrm{f}=0.33$ (data in Fig.~\ref{Varying cycle times}). This results from the increased thickness of the stratification layer without a corresponding increase in the thickness of the interpenetrating portion of the layer (as noted with respect to Fig.~\ref{Varying cycle times}) when the duration $t_j$ is too long compared to $\tau$. The consequence is that the segregation index increases with increasing $t_j$, even though the length of the penetrating layer is unchanged [Fig. ~\ref{fig:penetration}(a)]. 

\section{Conclusions}
	
We have described a continuum-model-based approach for modeling stratification of granular material in unsteady bounded heap flow, which can potentially serve as a prototype for modeling segregation in other transient and periodically varying flows. Specifically, by using the unsteady form of an advection-diffusion-segregation equation and accounting for the propagation of the moving front of particles down the slope of the heap after a change in the flow rate, we demonstrate accurate modeling of the dynamics of stratified layer formation. 

This model allows us to explore the potential for a modulated feed rate in heap flows of bidisperse particle mixtures to intentionally create layers of the two species so that the material is effectively mixed at length scales greater than the combined layer thickness. This layered configuration can have advantages over the usual fully segregated pattern that occurs for constant flow rates with the same average flow. For instance, if a stratified heap is formed using flow modulation when filling a hopper, the layers will remix as the hopper is discharged~\cite{Xiao2017}, whereas a fully segregated heap formed when filling a hopper at a constant fill rate will result in an excess of small particles at the beginning of the discharge and an excess of large particles at the end of the discharge~\cite{xiao2019continuum, Deng2020}.

With regard to heap segregation, further work is necessary to fully understand the impact of the feed flow parameters (rates and durations) to produce optimal stratification. The flow kinematics model used here (Figs.~\ref{Slow_to_fast} and \ref{fast_to_slow}) only considers a single isolated transition between high and low feed rates or vice versa. More work is necessary to model and understand flow rate modulation in which multiple transitions occur simultaneously in the heap, i.e.\ the feed rate is changed before the wedge front from the previous change reaches the downstream endwall, or in which the feed rate is varied continuously.   Furthermore, this approach could be extended to density-bidisperse, polydisperse, and combined size- and density-bidisperse mixtures as well as non-spherical particles, since these can all be modeled using the advection-diffusion-segregation approach (Eq.~\ref{bidisperse_eqn}) used here~\cite{Deng2020, Duan2021, HongyiDensity, Zhao2017, Jones2021}.

More generally, we have not considered flow modulation in 3D geometries in which the influence of sidewall friction on flow kinematics is greatly reduced or eliminated and the transient flow structures differ from the simple wedge shape~\cite{Isner2020b, Isner2020a}. Modulated 3D flows would likely need to be described by different forms of Eqs.~\ref{front_position_eqn} and \ref{flow_rate_eqn}, and may result in different transient phenomena (including unstable transients~\cite{Altshuler2008}). Nevertheless, these results demonstrate the potential for modeling feed flow modulation and using it to intentionally generate stratified patterns in segregating granular flows. Furthermore, as is evident here, coupling particle segregation with the deposition and erosion of particles from the fixed bed is challenging from both physics and numerical modeling standpoints. However, we have demonstrated that the approach is feasible and potentially could be applied in many industrial and geophysical flow situations.

\section*{Acknowledgement}
This material is based upon work supported by the National Science Foundation under Grant No. CBET-1511450. Authors HX and ZD contributed equally to this research. We are grateful for useful discussions with Dr. John Hecht and Dr. Yi Fan.





%\bibliographystyle{plainnat}
\bibliographystyle{elsarticle-num} 
%\bibliography{Citation}

\end{document}